\documentclass[twocolumn,showpacs,amsmath,amssymb,prb]{revtex4}

\usepackage{graphicx}
\usepackage{dcolumn}
\usepackage{bm}

\begin{document}

\preprint{APS/123-QED}
\title{Temperature-field phase diagram of geometrically frustrated CePdAl}

\author{$^{1}$Hengcan Zhao}
\author{$^{1}$Jiahao Zhang}
\author{$^{1}$Sile Hu}
\author{$^2$Yosikazu Isikawa}
\author{$^{1}$Jianlin Luo}
\author{$^{1,3}$Frank Steglich}
\author{$^{1}$Peijie Sun}
 \email{pjsun@iphy.ac.cn}
\affiliation{%
$^1$Beijing National Laboratory for Condensed Matter Physics, Institute of Physics, Chinese Academy of Sciences, Beijing 100190, China \\
$^2$Graduate School of Science and Engineering, University of Toyama, Toyama 930-8555, Japan\\
$^3$Max Planck Institute for Chemical Physics of Solids, 
D-01187 Dresden, Germany
}%

\date{\today}

\begin{abstract}
From the electrical resistivity and ac susceptibility measurements down to $T$ = 40 mK, a rich temperature-field phase diagram has been constructed for the geometrically frustrated heavy-fermion compound CePdAl. In the limit of zero temperature, the field-induced suppression of the antiferromagnetism in this compound results in a sequence of phase transitions/corssovers, rather than a single critical point, in a critical region of field 3$-$5 T. In the lowest temperature region, a power-law description of the temperature-dependent resistivity yields $A$$\cdot$$T^n$ with $n$$>$2 for all applied fields. However, a quadratic-in-temperature resistivity term can be extracted when assuming an additional electron scattering channel from magnon-like excitations, with $A$ diverging upon approaching the critical regime. Our observations suggest CePdAl to be a new playground for exotic physics arising from the competition between Kondo screening and geometrical frustration.
\end{abstract}

\pacs{Valid PACS appear here}
\maketitle

While identifying the magnetic ground state of a geometrically frustrated insulating spin system continues to be an intriguing topic in condensed matter physics \cite{Balents}, frustrated metallic magnets with Kondo screening emerge as a new playground for exotic physics \cite{qimiao1,coleman, qimiao, aronson}. The ground state of a Kondo lattice composed of localized $f$ electrons and a surrounding Fermi sea in the absence of geometrical frustration has been well described by Doniach's phase diagram \cite{doniach}. At one side of an appropriate external tuning parameter such as pressure, the indirect exchange interaction between $f$ sites, i.e., the Ruderman-Kittel-Kasuya-Yosida (RKKY) interaction tends to drive the system to be magnetically, usually antiferromagnetically, ordered. At the other, the Kondo effect prevails and screens the magnetic moments of $f$ electrons, giving rise to a Landau Fermi liquid (LFL) with enhanced mass of composite quasiparticles. If the magnetic order disappears in a smooth way as a function of the tuning parameter, a quantum critical point (QCP) ensues. In the vicinity of a QCP where quantum fluctuations are dominating over thermal fluctuations, interesting physics like unconventional superconductivity and non-Fermi liquid (nFL) phenomena may arise \cite{qcp}. The presence of geometrical frustration may lead to additional quantum fluctuations of local moments, adding to the delicate competition between the Kondo effect and RKKY interaction. Theoretically, a global phase diagram which goes beyond the single QCP in Doniach's phase diagram has been proposed for such a system, with quantum spin liquid, magnetic order, LFL and nFL behavior in different parts of phase space \cite{qimiao,coleman}. Experimentally, however, this topic is largely unexplored,
partly due to the lack of appropriate frustrated Kondo systems.

A Kondo-lattice series coming into attention in this context crystallizes in the hexagonal ZrNiAl-type structure with space group $P$\=62$m$ \cite{YbAgGe,fritsch}. There, the rare-earth atoms occupying the Zr site form a quasi-Kagome lattice with corner-sharing equilateral triangles in the $ab$-plane \cite{kagome}. Geometrical frustration has been discussed as having a profound influence to the ground-state physics of this compound series. For example, a complex temperature-field ($T$-$B$) phase diagram was found in a Yb-based homologue, YbAgGe, with multiple magnetic phases and a nFL regime that was ascribed to quantum bicriticality in the presence of significant frustration \cite{YbAgGe,tokiwa0}. Moreover, it was demonstrated for CeRhSn by thermal expansion measurements that the geometrical frustration in the $ab$ plane is responsible for its zero-field quantum criticality \cite{tokiwa}. CePdAl of this series is a typical heavy-fermion compound, with a single antiferromagnetic (AFM) transition at $T_{\rm N}$ = 2.7 K and a Sommerfeld coefficient of the electronic specific heat, $\gamma$ = 270 mJ/mol K$^2$ \cite{kitazawa,isikawa,isikawa2}. In this compound, the intra-plane nearest neighbour distance of Ce is 3.722 \AA, significantly shorter than the inter-plane distance 4.233 \AA \cite{fritsch}. This guarantees a significant 2-dimensional geometrical frustration in the basal plane. Consequently, neutron scattering measurements have detected that the antiferromagnetism below $T_{\rm N}$ involves only two-thirds of the Ce moments, with an incommensurate propagation vector $\textbf{Q}$ = [1/2, 0, 0.35] \cite{donni}. One-third of the Ce moments remain disordered below $T_{\rm N}$, down to at least 30 mK \cite{nmr}.  In other words, the equivalent (3$f$) sites of Ce atoms in paramagnetic regime break down into two inequivalent groups at $T$ $<$ $T_{\rm N}$:  Ce(1) and Ce(3) which are involved in the AFM ordering and Ce(2) that forms a frustrated network hindering the magnetic interaction between Ce(1) and Ce(3) \cite{donni}. Noticeably, the strong uniaxial crystal electric field (CEF) leads CePdAl to be an effective spin 1/2 Ising-like system \cite{isikawa}, with a susceptibility ratio $\chi_{c}$/$\chi_{ab}$ $>$ 20 at $T$ = 2 K. This is to be compared to the homologue YbAgGe, which has its magnetic easy direction in the basal plane and only a moderate anisotropy, $\chi_{ab}$/$\chi_{c}$ $\approx$ 3 \cite{katoh}.

An unusual AFM state in CePdAl has also been recognized in thermodynamic and magnetic measurements. While specific heat shows a well-defined $\lambda$-shaped anomaly at 2.7 K \cite{fritsch,isikawa2} signalling the onset of the AFM ordering, magnetic susceptibility $\chi(T)$ exhibits a broad maximum at higher temperature, $T$ $\approx$ 4 K \cite{isikawa}.
This discrepancy has been discussed in terms of enhanced short-range magnetic correlations \cite{nmr}.
We note that, this phenomenon appears to be common in many frustrated magnets. In such systems, $T_{\rm N}$, if it exists, may lead to a kink in $\chi(T)$ below the temperature of a broad maximum associated with enhanced magnetic fluctuations, as is seen, for example, in the honeycomb lattice iridates Na$_2$IrO$_3$ and Li$_2$IrO$_3$ \cite{iridate}. Furthermrore, a spin-glass-like nature of the ordered state has been detected by the frequency dependence of the ac susceptibility for a polycrystalline sample of CePdAl \cite{li06}.  Hydrostatic pressure \cite{goto} and chemical substitution \cite{fritsch} have been shown to be able to suppress the AFM ordering and establish a QCP. However, the specific nature of this QCP, influenced by the enhanced quantum fluctuations associated with geometrical frustration, remains yet to be explored. In this work, we use magnetic field as tuning parameter to explore the phase diagram of CePdAl. By measuring the resistivity and ac susceptibility on high quality single crystals down to $T$ = 40 mK and up to $B$ = 14 T, we were able to construct a detailed $T$-$B$ phase diagram. Extrapolating our data to $T$ = 0 K we conclude that multiple quantum phase transitions, rather than a single QCP, emerge consecutively in a confined critical region (3$-$5 T), following a smooth field-suppression of the AFM ordering. Fitting the low-temperature resistivity to a power-law function results in an exponent $n$ $>$ 2 for all applied fields; in other words, we don't observed either nFL behavior with $n$ $<$ 2 or Fermi liquid behavior with $n$ = 2. However, the existence of a field-induced QCP is indeed suggested by the diverging $A$ coefficients estimated from the extracted quadratic-in-temperature resistivity term, which was obtained after taking into account a scattering channel from magnon-like excitations.

A high-quality single crystal of CePdAl was grown by the Czochralski technique in an argon gas atmosphere with induction heating \cite{isikawa}.  The temperature and field-dependent electrical resistivity $\rho(T, B)$ was measured with a four-probe technique down to $T$ = 40 mK using a $^3$He-$^4$He dilution refrigerator in a magnetic field up to 14 T. Ac susceptibility $\chi_{\rm ac}(T, B)$ was measured in the same temperature and field ranges, utilizing a lock-in amplifier. The magnetic field was applied along the crystallographic $c$ axis, which is the direction of the Ising-like moment, in all the measurements. This allows for a suppression of the AFM ordering in a moderate field less than 5 T. By contrast, if the field is oriented within the $ab$-plane, we estimate the critical field to be larger than 30 T, far beyond the fields accessible in our measurements.

\begin{figure}[t]
\includegraphics[width=0.96\linewidth]{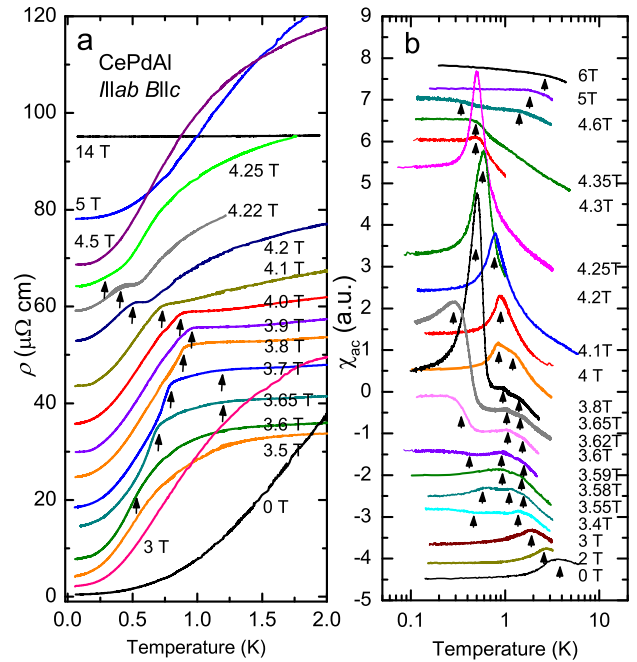}
\caption{Electrical resistivity $\rho(T)$ (a) and ac susceptibility $\chi_{\rm ac}(T)$ (b) as a function of temperature measured in various dc magnetic fields applied along the $c$ axis of CePdAl. The values of $\rho(T)$ are offset vertically for clarity. Arrows indicate possible phase transitions or crossovers.
\label{nus.eps}}
\end{figure}

Figure 1a shows the in-plane electrical resistivity $\rho(T)$ for $T$ $<$ 2 K measured in various magnetic fields $B$$\parallel$$c$. Previous reports \cite{isikawa,kitazawa} have shown that $\rho(T)$ at zero field exhibits a broad maximum at $T$ $\approx$ 4 K, while the AFM ordering at $T_{\rm N}$ = 2.7 K can be hardly detected from resistivity.
Increasing the magnetic field to above 3 T, a clear anomaly appears to be visible in the $\rho(T)$ curves, as indicated by the vertical arrows in Fig. 1a. With further increasing field, the position of the anomaly initially increases, then decreases for $B$ $>$ 3.9 T, forming a distinct dome in the $T$-$B$ phase diagram to be discussed below. In certain fields, e.g., $B$ $=$ 3.6 and 3.7 T, more than one anomaly can even be confirmed. Similarly, all these anomalies are also seen in the ac susceptibility $\chi_{\rm ac}(T)$ measured at various superimposed dc fields, as shown in  Fig. 1b. As $\chi_{\rm ac}$ is more sensitive to phase transitions, it is possible to trace the smooth suppress of $T_{\rm N}$ at $B$ $<$ 3 T in ac susceptibility measurements. Likewise, the development of multiple phase boundaries in the critical regime between $B$ = 3 and 5 T is also clearly revealed by these measurements.

\begin{figure}[t]
\includegraphics[width=0.95\linewidth]{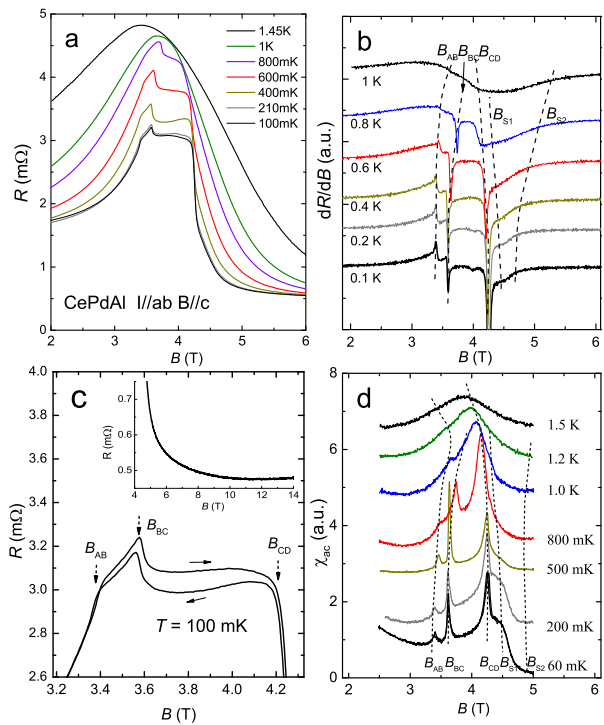}
\caption{Low-temperature isotherms of resistivity and ac susceptibility measured as a function of field. (a) $\rho(B)$ measured in the configuration $I$$\parallel$$ab$ and $B$$\parallel$$c$. (b) Derivative of the resistivity with respect to field, d$\rho$/d$B$, to highlight the phase boundaries. (c) A round-trip field scan of resistivity at $T$ = 100 mK to illustrate the hysteresis of the multiple metamagnetic transitions. Inset shows the profile of $\rho(B)$ in the higher field region up to 14 T. (d) ac susceptibility $\chi_{\rm ac}(B)$ measured with a superimposed dc field applied along $c$. Dashed lines in panels b and d, labeled as $B_{\rm AB}$, $B_{\rm BC}$, $B_{\rm CD}$, $B_{\rm S1}$ and $B_{\rm S2}$, denote probable phase boundaries. See text for details.
\label{nus.eps}}
\end{figure}

The phase boundaries seen in the $\rho(T)$ and $\chi_{\rm ac}(T)$ curves are further confirmed by isothermal measurements as a function of field. As shown in Fig. 2a, $\rho(B)$ at $T$ $<$ 1 K is a complicated function of $B$ and displays a $\pi$-shaped field-variation with multiple sharp features in between $B$ = 3 and 5 T. This is in good agreement to previous $\rho(B)$ measurements performed at a single temperature, $T$ = 0.6 K \cite{goto}. Similar $\rho(B)$ profiles have been observed in, for example, CeAuSb$_2$ \cite{Au112} and CeAgBi$_2$ \cite{Ag112}, whose zero-field AFM states evolve into multiple metamagnetic states upon increasing the field. The sequence of the multiple phase transitions in CePdAl is much more clearly evidenced by corresponding jumps and shoulders in the field dependence of d$\rho(B)$/d$B$, cf. Fig. 2b. The positions of the three sharp jumps are denoted as $B_{\rm AB}$, $B_{\rm BC}$, and $B_{\rm CD}$, and those of the two following shoulders, as $B_{\rm S1}$ and $B_{\rm S2}$, from the low to the high field side. A careful field scan for $\rho(B)$ at temperatures below $T$ $\approx$ 600 mK (Fig. 2c) reveals a weak hysteresis for $B_{\rm AB}$, $B_{\rm BC}$, and $B_{\rm CD}$, hinting at first-order phase transitions. Likewise, similar phase boundaries as revealed in d$\rho(B)$/d$B$ are unambiguously detected in the $\chi_{\rm ac}(B)$ isotherms measured at different temperatures below 1.5 K (cf. Fig. 2d). Note that the positions of these anomalies agree well to the field-induced metamagnetic transitions observed in the magnetization curve measured at $T$ = 0.51 K \cite{goto}. At $T$ $>$ 0.8 K, the sharp features at the critical fields $B_{\rm AB}$, $B_{\rm BC}$, and $B_{\rm CD}$ as seen in $\rho(B)$ and $\chi_{\rm ac}(B)$ apparently broaden suggesting second-order phase transitions at these elevated temperatures.

As shown in Fig. 3, the $T$-$B$ phase diagram constructed from the aforementioned measurements consists of a variety of phase lines, separating at least six different low-temperature phases (A-E). As already mentioned, a smooth suppression of the AFM order by field at $B$ $<$ 3 T does no lead to a single QCP as observed in prototypical heavy-fermion systems, e.g., YbRh$_2$Si$_2$ \cite{YRS}. In parallel to the case of YbAgGe \cite{YbAgGe,tokiwa0}, this is believed to be a consequence of geometrical frustration of local magnetic moments forming the quasi-Kagome lattice. The sequence of the multiple phase boundaries is observed in a very confined critical phase space in the vicinity of a magnetic dome (phase C), i.e., between $B$ = 3 and 5 T. This is to be compared to the $T$-$B$ phase diagram composed of a similar sequence of transitions in, e.g., YbAgGe \cite{YbAgGe} and CeAgBi$_2$ \cite{Ag112}, where the metamagnetic transitions are spread over a considerably large phase space. Following previous neutron scattering experiments \cite{donni}, the large phase space at the low-temperature, low-field corner (phase A) is an incommensurate AFM phase associated with two-thirds Ce moments at sites Ce(1) and Ce(3), though the incommensurate component of the wave vector has been found to be slightly field dependent \cite{prokes}. In view of the AFM nature and the strong magnetic anisotropy in phase A, it is tempted to assume that the phase boundaries at $B_{\rm AB}$ and $B_{\rm BC}$ are due to spin flop of the two-thirds ordered moments and phases B and C are canted AFM phases. However, the situation seems not that simple. Neutron diffraction measurements performed on single crystals indicate that the geometrical frustration is partially lifted in field and the moments at Ce(2) sites participate, at least partially, in the field-induced magnetic phases \cite{prokes}.  At a first glance, phase D seems a high-field extension of phase B, forming a single phase that encircles the smaller dome of phase C. This is very likely not the case: As evidenced from the low-temperature $\rho(B)$ curves, the magnetic fluctuations that dominate electron scattering in phase B and D are significantly different.

\begin{figure}[t]
\includegraphics[width=0.95\linewidth]{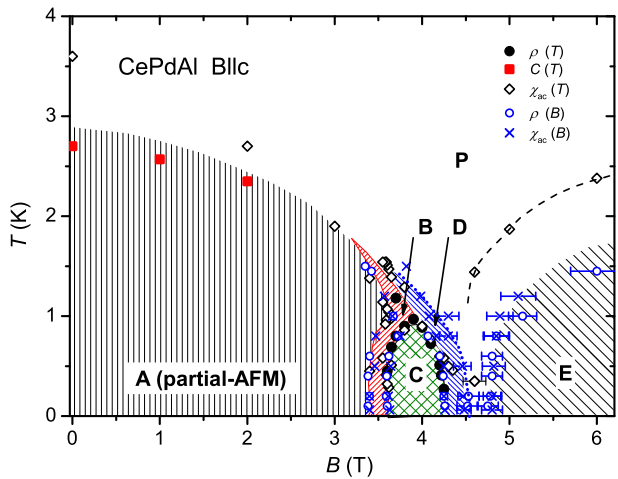}
\caption{$T$-$B$ phase diagram for CePdAl constructed from results of resistivity and ac susceptibility measurements as a function of both temperature and field. $T_{\rm N}$ obtained from specific-heat measurements is also shown. Note the critical field region between $B$=3 and 5 T, where a sequence of phase boundaries were detected. As the phase boundary of the high-field phase E is derived from a shoulder at $B_{\rm S2}$ in d$\rho(B)$/d$B$ and $\chi_{\rm ac}$($B$), it has a large error bar. The dashed line connecting several data points at $B$$>$4.5 T, which were extracted from $\chi_{\rm ac}(T)$ measurement, may reflect the broad crossover between phase P and phase E or, alternatively, a Kondo breakdown line.
\label{nus.eps}}
\end{figure}

Compared to the apparent magnetic nature of phases B and C, which shows significant impact of spin-fluctuations on the $T$/$B$-dependence of the resistivity, the nature of phase D, where the spins are substantially aligned and the low-$T$ magnetoresistance (Fig. 2a) is abruptly diminished, remains elusive.
As seen in Fig. 2b and 2d, the phase boundary between the paramagnetic P phase and phase D is marked by a broad shoulder at $B_{\rm S1}$ in $\rho(B)$ and $\chi_{\rm ac}(B)$, with a large error bar along the field direction, cf. Fig. 3. Therefore, the $B_{\rm S1}$ line may simply reflect a crossover. Here, one possibility is to refer phase D as a transitional one between the magnetically ordered phase C and paramagnetic phase P. In this case, an exciting scenario would be a field-induced metallic spin liquid state, the magnetic fluctuations associated with it giving rise to the shoulder in both d$\rho$/d$B$($B$) (Fig. 2b) and $\chi_{\rm ac}$$(B)$ (Fig. 2d). Such a phase has been proposed for 6\% Ir substituted YbRh$_2$Si$_2$ in connection with a Kondo breakdown detached from its AFM QCP \cite{svan09}.

The nature of the presumably nonmagnetic phase E is elusive, too. In Doniach's phase diagram, this would represent a LFL state with a large Fermi surface due to heavy quasiparticles. Characterizing the coherent propagation of Kondo singlets, a positive magnetoresistance as found in most nonmagnetic metals, is usually observed in this phase, as holds for YbRh$_2$Si$_2$ \cite{svan}. However, for CePdAl, the magnetoresistance measured at $T$ = 100 mK is negative up to at least 12 T (cf. Fig. 2c inset), hinting at significant spin fluctuations which persist deep inside the E phase.  
As the AFM ordering is already suppressed in phase E, one expects that the inequivalency of ordered and disordered Ce sites is lifted. In this case, how zero-point fluctuation arising from geometrical frustration can compete Kondo screening in forming an unconventional Kondo singlet remains an extremely interesting and challenging topic.
In Fig. 3, a dashed line connecting several data points obtained from the position of the shoulder in $\chi_{\rm ac}(T)$ is shown at $B$ $>$ 4.5 T. As is known for YbRh$_2$Si$_2$ \cite{svan, svan09}, in the regime of quantum criticality, a peak has been detected in ac susceptibility as a function of temperature, reflecting the energy scale of local Kondo breakdown. More detailed experiments are needed in order to clarify  whether this can take place also in CePdAl. Alternatively, this line may simply reflect the broad crossover between phase P and phase E.

\begin{figure}
\includegraphics[width=0.9\linewidth]{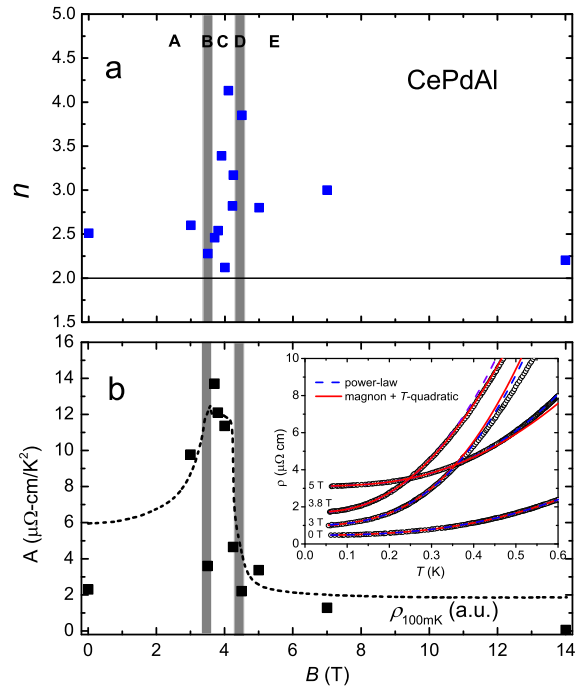}
\caption{(a) Magnetic field dependence of the exponent $n$ estimated by fitting the resistivity in the lowest temperature region (typically below 300 mK) to $\rho$ = $\rho_0$ + $A$$\cdot$$T^n$. (b) $A$ coefficient of the extracted quadratic-in-temperature resistivity term from the raw data. See text for detail. The resistivity measured at $T$ = 100 mK, regarded as the residual resistivity $\rho_0$, is shown by the dotted line for comparison. The thick vertical lines are used to distinguish regions of different phases.  Inset: The power-law fits as well as the fits taking into account scattering of magnon-like excitations to low-temperature $\rho(T)$ curves (offset vertically) measured in several typical fields.
\label{nus.eps}}
\end{figure}

Given the sequence of first-order metamagnetic phase transitions and the following continuous phase boundaries that are approaching to zero temperature in a confined field regime, it is interesting to explore the $T$-dependence of $\rho(T)$ and search for deviations from the LFL-type $T^2$ behavior due to dominating quasiparticle-quasiparticle scattering. In fact, in YbAgGe \cite{nik}, a linear $\rho(T)$ has been observed in an extended field region beyond the suppression of AFM state. As displayed in Fig. 4a and the inset of Fig. 4b, our fitting of the measured $\rho(T)$ below 300 mK to the power-law expression $\rho$ = $\rho_0$ + A$T^n$ yields an exponent $n$ $>$ 2 in the whole field region, with a large scattering in the vicinity of the critical fields (3$-$5 T). The absence of $n$ $<$ 2, a fingerprint for nFL behavior, is either intrinsic, considering the first-order nature of the metamagnetic phase transitions, or masked by other inherent magnetic excitations. At least for the magnetic phases A-C, the quasiparticles from AFM magnons will add to the quasiparticle-quasiparticle scattering, giving rise to $n$ $>$ 2. The fact that $n$ $>$ 2 in the whole field range including the high-field E phase indicates some kind of robust spin-dependent scattering of the quasiparticles. This is consistent with the negative magnetoresistance persisting deep into the phase E (cf. inset of Fig. 2c). A straightforward interpretation, as mentioned above, is that the geometrical frustration hinders the formation of a nonmagnetic coherent heavy-fermion state, giving rise to a strange metal regime even in the phase E \cite{coleman}. To further investigate the possibility of an underlying QCP masked by yet unidentified magnetic excitations, we tried to fit all $\rho(T)$ curves assuming an extra scattering channel due to magnon-like excitations, in addition to a quadratic dependence \cite{magnon}: $\rho(T)$ = $\rho_0$ + $a$$T$(1+2$T$/$\Delta_s$)exp(-$\Delta_s$/$T$)+$A$$\cdot$$T^2$. Here $\Delta_s$ is the energy gap for magnon excitations. Indeed, we can describe most of the $\rho(T)$ curves in the lowest temperature region reasonably well, with a tiny $\Delta_s$ $<$ 1 K, cf. inset of Fig. 4b. The obtained coefficient $A$ is plotted in Fig. 4b. An apparent divergence is observed in the field interval 3 $-$ 5 T, despite considerable scattering of the data points. The values of $\rho(B)$ measured at $T$ = 100 mK, regarded as residual resistance $\rho_0$, is re-plotted in Fig. 4b, too. There is a good agreement between the profiles of $A(B)$ and $\rho_0(B)$, as have been seen in, e.g., Sr$_3$Ru$_2$O$_7$ \cite{Ru327}. Similarly, the latter compound has an unconventional QCP approached by fully suppressing the critical end point of first-order metamagnetic phase transition line by applying a magnetic field.  It is interesting to note that in Sr$_3$Ru$_2$O$_7$, a power-law description of the low temperature $\rho(T)$ curves, with $n$ $>$ 2, was also possible, though in the vicinity of its QCP only.

In conclusion, we have constructed a detailed $T$-$B$ phase diagram for CePdAl, a heavy-fermion compound with geometrically frustrated Ce moments in the Kagome-like $ab$ plane. When a magnetic field is applied along the magnetic easy $c$-direction, $T_{\rm N}$ is smoothly depressed up to $B$ = 3 T. With further increasing field, instead of a single QCP, a critical regime with multiple phase transitions/crossovers are observed in a confined field region, 3$-$5 T. Over the entire field regime and down to the lowest temperature, no apparent nFL behavior, but rather a power-law dependence with an enhanced exponent $n$ $>$ 2 was obtained for $\rho(T)$.  Only after taking into account scattering from additional magnon-like excitations, one can extract a quadratic-in-temperature term of resistivity, with diverging $A$ coefficient towards the critical field region, suggesting an underlying putative QCP. Even more, between the magnetic phase C and the paramagnetic phase P, there appears a transitional phase D which deserves special attention in view of potential metallic spin liquid state. Similarly, phase E may be unconventional, too, in view of pronounced magnetic fluctuations. Further in-depth investigations are badly needed in order to get a deeper insight into the manifold of potentially novel ground states.

The authors are grateful to discussion with A. Strydom, Q. Si, R. Yu, P. Gegenwart, Y. Tokiwa, J-G. Cheng and Y.F. Yang. This work was supported by the MOST of China (Grant No: 2015CB921303), the National Science Foundation of China (Grant No:11474332), and the Chinese Academy of Sciences through the strategic priority research program (XDB07020200).


\begin{thebibliography}{99}
\bibitem{kagome} It is refered to as a quasi-Kagome lattice because it is formed by corner-sharing equilateral triangles as in a Kagome lattice. However, in this structure every rare earth atom has only two nearest neighbers, not three as in a prototype Kagome lattice.
\end{thebibliography}
\end{document}